\documentclass[twocolumn,a4paper]{article}

\usepackage{graphicx} 
\usepackage{siunitx}

\usepackage{helvet}

\usepackage[utf8]{inputenc}

\usepackage[font=footnotesize, skip = 1pt, labelfont=bf]{caption}
\usepackage[font=footnotesize]{subcaption}

\usepackage{dcolumn}           
\newcolumntype{.}   {D{.}{.}{-1}} 
\newcolumntype{d}[1]{D{.}{.}{#1}} 
\newcolumntype{e}   {D{E}{E}{-1}} 
\newcolumntype{E}[1]{D{E}{E}{#1}} 

\usepackage{amsmath}
\DeclareMathSizes{7}{7}{3}{3} 
\usepackage{pifont}
\usepackage{amsfonts}
\usepackage{amssymb}
\usepackage{amsthm}

\usepackage{float}

\usepackage{sectsty}
\newcommand{\myFontSize}{\fontsize{9}{0}\selectfont}
\sectionfont{\myFontSize}       
\subsectionfont{\myFontSize} 

\usepackage{titlesec}
\usepackage{booktabs}
\usepackage{csquotes}

\titlespacing*{\section}{0pt}{10pt}{0pt}
\titlespacing*{\subsection}{0pt}{10pt}{0pt}

\usepackage{secdot}
\usepackage{epstopdf}
\sectiondot{subsection}
\sectionpunct{section}{. }    
\sectionpunct{subsection}{. } 

\usepackage{hyperref} 

\title{\bfseries A Bayesian Application in Judicial Decisions \\ \small{Extended Abstract}}
\date{\texttt{2019-12-22}}
\author{Filipe J. Zabala \\ \small{\texttt{filipe.zabala@pucrs.br}} \\ \\ 
\href{http://www.pucrs.br/en}{PUCRS} \\ \small{Pontifícia Universidade Católica do Rio Grande do Sul}}

\begin{document}

%
\twocolumn[
\begin{@twocolumnfalse}
  \maketitle

\begin{abstract}

This paper presents a new tool to support the decision concerning moral damage indemnity values of the judiciary of Rio Grande do Sul, Brazil. A Bayesian approach is given, in order to allow the assignment of the magistrate's opinion about such indemnity amounts, based on historical values. The solution is delivered in free software using public data, in order to permit future audits. 

\noindent{{\bf Keywords:}} Indemnity amounts, moral damage, jurimetrics, bayesian. \\

\end{abstract}

\end{@twocolumnfalse}
]
	\section{Introduction}
\label{sec:intro}

Assigning moral damage indemnity values is an open problem in the judiciary. In partnership with TJRS\footnote{Tribunal de Justiça do Rio Grande do Sul.} judges, PUCRS professors and students, a tool is being structured to support the magistrate's decision in cases of moral damage. The projected schedule is from January 2020, with expected duration of eight months. The team is composed by statisticians, lawyers, computer scientists and scholarship students. The theoretical framework is being structured to ensure the best possible use of available data, facilitating end-user operation. 

Such a tool is being developed under the Bayesian paradigm, where the magistrate must indicate his opinion about the value of the cause during the evaluation of the judicial process data. The bayesian approach presents tools to address solutions involving \textit{small} to \textit{big} data structures, using available datasets to update the decision maker's opinion. In \cite{kadane2012probability} is presented a lecture given in the XI Brazilian Meeting on Bayesian Statistics, concerning the case of Kansas cellphone users\footnote{Quin Jackson et al. v. Sprint Nextel Corporation, Case No. 09-cv-2192 (N.D. Ill.).}, in which `Sprint-Nextel was sued for conspiring with other cell phone providers to impose high prices for text-messaging'. Joseph Kadane discusses the probability sampling concept and its application, pointing out that in the present case `classical statistics did not address the court’s question, but Bayesian analysis did'. The same happens in cases involving the judiciary, in the sense that is needed a formal structure to elicit the expert's opinion and update it. The classical approach does not permit such structure, with no formal alternative to deal with the subjectivity inherent to the decision making process. According to Bruno de Finetti, \textit{(t)here is no way, however, in which the individual can avoid the burden of responsibility for his own evaluations. The key cannot be found that will unlock the enchanted garden wherein, among the fairy-rings and the shrubs of magic wands, beneath the trees laden with monads and noumena, blossom forth the flowers of `Probabilitas realis'. With these fabulous blooms safely in our button-holes we would be spared the necessity of forming opinions, and the heavy loads we bear upon our necks would be rendered superfluous once and for all.} \cite[p. 42]{finetti1975theory}

	\section{Background}
\label{sec: backg}

The historical framework for this purpose can be traced since \cite{leibniz1666dissertatio}, considering also the works of \cite{bernoulli1709dissertatio},  \cite{bernoulli1713ars}, \cite{bayes1763essay}, \cite{laplace1774memoire} and \cite{condorcet1785essay}. From the 19th century it can be considered the works of \cite{guerry1833essai}, \cite{poisson1837recherches}, \cite{holmes1881common} and \cite{holmes1897path}. These ancient authors inspired Legal Realists as well Lee Loevinger \cite{loevinger1949jurimetrics}, that in 1949 wrote a manifesto in defense of rationality in the law. For him, \textit{(l)awyers gather data, which they call `evidence'; scientists gather evidence, which they call `data'. Both terms mean the same thing, which is intellectual support for some conclusion or proposition.} \cite[p. 323]{loevinger1992standards}

Other modern researchers considers the close proximity between science and law. \textit{The linguistic style in which legislation is normally written has many similarities with the language of logic programming. (...) These extensions include the introduction of types, relative clauses, both ordinary negation and negation by failure, integrity constraints, metalevel reasoning and procedural notation.} \cite[p. 325]{kowalski1995legislation}

The use of quantitative methods in law is therefore longstanding. In Loevinger's words, \textit{(t)he branch of mathematics that appears to be of the most immediate practical utility in the fields of law and the behavioral sciences is statistics. There is much in statistics that is of present practical application in day-to-day legal problems and it has good claim to be included in every law school curriculum.} \cite[p. 262]{loevinger1961jurimetrics}. In this sense the Pontifical Chatholic University of Rio Grande do Sul is starting a tradition in brazilian jurimetrics, that began with some 2012 lectures\footnote{\url{https://www.youtube.com/watch?v=kKipITR9iSM}}, outreach videos\footnote{\url{https://www.youtube.com/watch?v=4HXPES4GnO8}} and is formalized by \cite{zabala2014jurimetria}, \cite{aires2017norm}, \cite{aires2019automatic}, \cite{aires2019classification}, \cite{zabala2019jurimetrics} and other upcoming works.

	\section{Methodology}
\label{sec: Methodology}

The values suggested by the system are based on TJRS's decision history. This history is public and was obtained with a web scraper written in Python language from the TJRS website\footnote{\url{https://www.tjrs.jus.br/site}}. This procedure is supported by the Access to Information Act from 18 November 2011 (Act no. 12.527/11, \cite{brasil2011lei}) and its regulation from 16 May 2012 (Act no. 7.724/12, \cite{brasil2012decreto}). According the 2011 Act, \textit{(i)t is the duty of public bodies and entities to promote, regardless of requirements, the disclosure in a place that is easily accessible, within the scope of their competences, of information of collective or general interest produced or guarded by them. (...) In order to comply with the caput, public bodies and entities shall use all legitimate means and instruments at their disposal, and disclosure on official websites of the World Wide Web (Internet) is mandatory. (...) Sites (...) shall (...) meet (...) the following requirements: I - contain content search tools that allow access to information in an objective, transparent, clear and in easy to understand language; II - enable the recording of reports in various electronic formats, including open and non-proprietary, such as spreadsheets and text, in order to facilitate the analysis of information; III - enable automated access by external systems in open, structured and machine readable formats; IV - disclose in detail the formats used for structuring the information; V - guarantee the authenticity and integrity of the information available for access; VI - keep updated the information available for access.} \cite[Article 8]{brasil2011lei}. 

The raw database has around 6 million entries, with data about the judges, counties, reference dates, as well as the abstract entry (\textit{ementa}) and entire content (\textit{inteiro teor}) of the cases. The extraction of relevant data from entry and entire content is being built with REGEX (REGular EXpressions) language, programmed by Computer students. Working in partnership, Law students supports discussions on legislation and legal framework. All the work is supervised by professors and magistrates to ensure the involvement of the entire team. The results must be programmed in a HTML tool, in order the magistrates to update their views in the light of available data.

\begin{figure*}[!h]
  \centering \includegraphics[width=\textwidth]{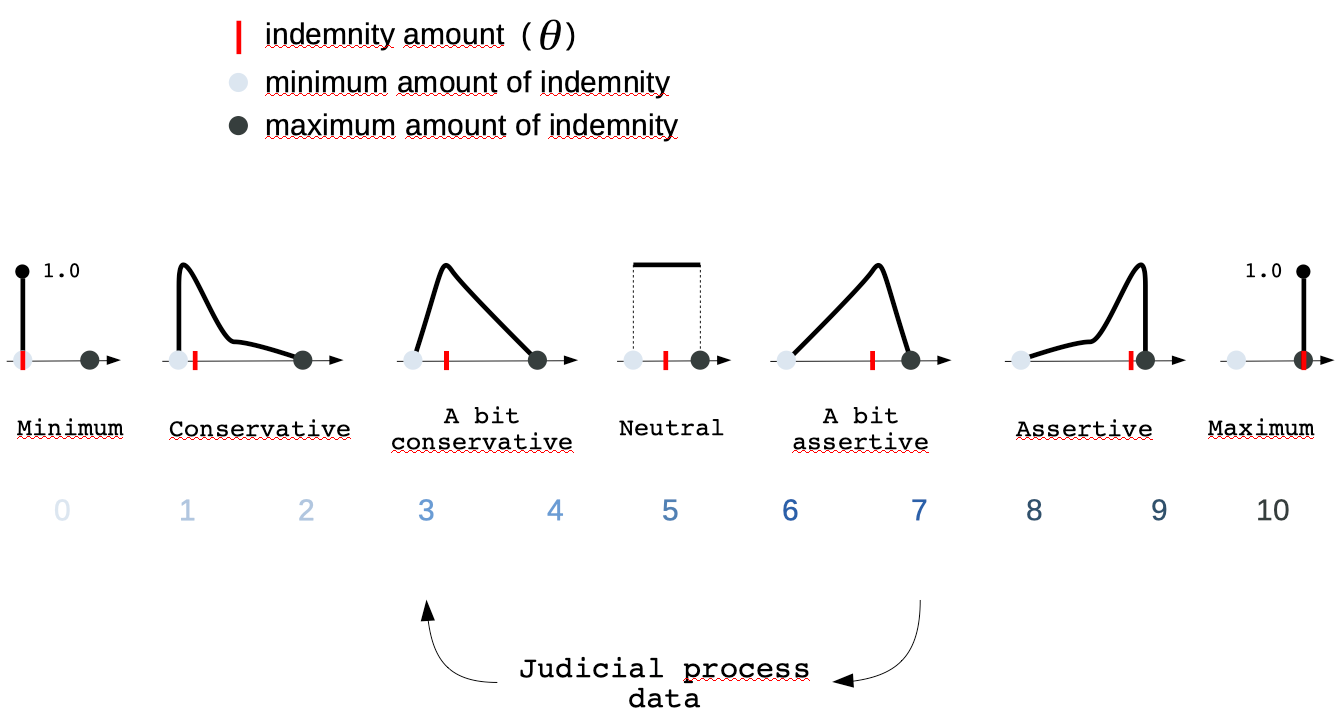}
  \caption{Working flow of indemnity amount estimation.}
  \label{fig1}
\end{figure*}

In Figure \ref{fig1} is shown a macro scheme for the proposed solution. The goal is to infer $\theta$, the indemnity amount for a particular case, based on the magistrate's opinion and the historical indemnity values. The judge is free to decide for the amount he deems most appropriate, and the system provides summary information to support the magistrate's decision. The opinion about how conservative (near the minimum $m$) os assertive (near to maximum $M$) is given in a ordinal scale from 0 (pay the minimum) to 10 (pay the maximum). To the levels of the ordinal scale are assigned parameters of a beta probability distribution, in order to model $\pi$, the proportion of historical range (the difference between the maximum and the minimum, anotated by $M-m$) must be paid. The symbolic representation is given by $\pi \sim Beta(\alpha,\beta)$, and the hyperparameters $\alpha$ and $\beta$ can be adjusted manually by the user. This approach is based in the technical report by \cite{wechsler2006analise}. The automatic definition of $\alpha$ and $\beta$ is given from historical summary measures as mean, variance and quantiles. After the definition about the hyperparameters -- based on the given ordinal scale or in its user-based definition -- the estimate of parameter $\theta$ is given by
\begin{equation}
\hat{\theta} = m + \hat{\pi} (M-m) + c,
\end{equation}

where $\hat{\pi}$ is the estimated value of the proportion of historical range to be paid, and $c$ is a constant user-definied that allows calibration on the suggested value and pre-defined as $c=0$. Since historical data are only a reference, the constant $c$ allows the user to define values that are below the historical minimum ($\hat{\pi}=0$ and $c<0$) or above the historical maximum ($\hat{\pi}=1$ and $c>0$).

The system must allow the filtering of the judicial processes by classes, dates, subjects, etc., in order to make the data closer to the considered case, according the system operator. The idea is to provide reference values to the decision maker. The report written in HTML may be linked to a code written in free software, that has the possibility to update the database and the involved (hyper)parameters.

	\section{Expected results}
\label{sec:resul}

Using the proposed tool concerning moral damage indemnity values on the court of TJRS, the following main results are expected:
\begin{itemize}
	\item data-driven decisions
	\item standardization of indemnity values
	\item reduction in discrepancy between similar cases
	\item increasing of the judicial security
\end{itemize}

As consequence of the main results, are expected also the extrapolation to other process types in TJRS, as well as the use of this and other data-driven methodologies in the brazilian judiciary. Because it is designed in open source software and supported by public databases, it may be the missing stimulus to make good use of available information in order to optimize the use of public resources, so poorly managed in recent decades.

 Finnaly, as Dennis V. Lindley points out on the foreword of \cite{finetti1974theory}, `we shall all be Bayesian by 2020'. By associating this philosophy with computationally efficient methods, it is possible to elucidate many issues in the Brazilian judiciary.


%
\bibliographystyle{abbrv}

\bibliography{references}

\end{document}